# Application of Global and One-Dimensional Local Optimization to Operating System Scheduler Tuning

George Anderson, Tshilidzi Marwala and Fulufhelo Vincent Nelwamondo

School of Electrical Engineering
University of Johannesburg
Johannesburg, South Africa
georgeganderson@gmail.com, tmarwala@uj.ac.za, fnelwamondo@csir.co.za

**Abstract**—This paper describes a study of comparison of global and one-dimensional local optimization methods to operating system scheduler tuning. The operating system scheduler we use is the Linux 2.6.23 Completely Fair Scheduler (CFS) running in simulator (LinSched). We have ported the Hackbench scheduler benchmark to this simulator and use this as the workload. The global optimization approach we use is Particle Swarm Optimization (PSO). We make use of Response Surface Methodology (RSM) to specify optimal parameters for our PSO implementation. The one-dimensional local optimization approach we use is the Golden Section method. In order to use this approach, we convert the scheduler tuning problem from one involving setting of three parameters to one involving the manipulation of one parameter. Our results show that the global optimization approach yields better response but the one- dimensional optimization approach converges to a solution faster than the global optimization approach.

## I. INTRODUCTION

Operating systems play the role of resource managers: modern computer systems consist of many resources such as main memory, hard disk drives and other storage devices, network interfaces, and the CPU [1]–[3]. Operating Systems consist of several components, each being responsible for managing a subset of the available resources. In this paper, we are concerned with optimizing the operating system scheduler by tuning its parameters. The job of the operating system scheduler is to decide which task should run on the CPU next, and for how long. Programs running on computer systems are represented as tasks, which keep track of allocated memory and various types of state. For a task to do its work, it must use the CPU i.e. the instructions that make up the task must be executed, or obeyed. A collection of these tasks running on a computer makes up a workload. Various metrics are used to assess the performance of a computer system. One is response time: operating system designers would like to make sure the time between when the user makes a request and it is handled is minimized. Another metric, and the one we are interested in, is turnaround (or completion) time. We want to minimize the time it takes a workload to finish its work. In this paper we study the Linux scheduler (known as the Completely Fair Scheduler) and apply global optimization (Particle Swarm Optimization, after meta-optimization) and local optimization (Golden Section method) to tuning it. Our results show that global optimization gives the best results, i.e. the smallest turnaround time, but local optimization gives faster convergence, with turnaround time that could also be acceptable. Our paper is organized as follows: Section II gives some background on research in optimization of software; Section III gives an overview of scheduling in the operating system, using Linux as an example; Section IV discusses operating system scheduler simulation; Section V discusses our experiments involving application of optimization methods to improving the performance of the Linux operating system process scheduler, involving both global optimization and local optimization; Section VI discusses our results; and Section VII has our conclusion.

## II. OPTIMIZATION IN SOFTWARE SYSTEMS

Anderson et al. [4] used a statistical optimization approach known as Response Surface Methodology (RSM) to tune them Linux operating system kernel. The RSM approach involved designing a limited number of experiments; each experiment involved setting each of three scheduling parameters. A benchmark is then run for each experiment. Based on the performance results, a model was obtained. This model was then optimized in order to give the best setting for each one of the three parameters. The authors achieved an 11% performance improvement when using the optimal parameter settings over when the Linux scheduler's default parameter settings were used. The approaches described in this paper shed new light on use of standard global and local optimization for tuning the operating system scheduler.

## III. OPERATING SYSTEM SCHEDULING IN LINUX

The job of the operating system scheduler is to decide which task should run on a CPU next and for how long [1]–[3]. Schedulers implement scheduling algorithms, which could have goals of minimizing metrics such as waiting time (the time tasks spend waiting to use a CPU), response time (the time between when an event occurs and it is handled), and maximizing metrics such as throughput (tasks completed per given period of time) and fairness (ensuring that each task gets a fair share of the CPU).

The Linux kernel, starting from version 2.6.23, makes use of a scheduler known as the Completely Fair Scheduler (CFS). Kumar [5] and Mauerer [6] describe the internals of the CFS. The Linux CFS scheduler keeps track of how much tasks are being treated unfairly. A task is being treated unfairly if it is not being executed by the CPU. It maintains a binary search tree (a red-black tree) that orders tasks according to how unfairly they are being treated. A task that has been treated the most unfairly (because it has been waiting to use the CPU for a long time) is selected to execute on the CPU at the next opportunity. The CFS scheduler is modular, making it readily extensible, and also supports group scheduling and scheduling classes.

## IV. OPERATING SYSTEM SCHEDULER SIMULATION

LinSched is a Linux scheduling simulator developed by Calandrino et al. [7]. The code related to the Linux scheduler was extracted from the Linux 2.6.23 kernel and modified, so that it runs as an ordinary process. The process also consists of a simulation-management component, which runs in a loop, until the maximum number of ticks has expired. When a task (process or thread) is running on a regular Linux kernel, each time it is scheduled, i.e. assigned to a CPU, its instructions are executed, continuing where they left of last time it was scheduled. After the process has acquired the CPU for a long-enough combined period, it exits. With LinSched, a function is declared to represent a particular type of task. Every time the task is assigned to use a

CPU, the function is called.

LinSched has many advantages for use as a research tool, over a kernel. One is that the scheduler can be modified and tested easily. A bug in the scheduling code caused only the LinSched process to crash. Working with an actual kernel requires more time because the kernel has to be compiled then rebooted. A bug could bring down the kernel together with all the other processes that are running. Then the cycle of rebooting to a trusted kernel, fixing the bug, and rebooting again continues. LinSched also allows fast validation of scheduling algorithms, since the scheduling sub-system is isolated. In our research we have chosen to use the LinSched simulator because it is the only one that is based directly on actual Linux code, while completely isolating the scheduling subsystem. Other simulators, e.g. AKULA [8], require the Linux scheduler's code to be ported to them.

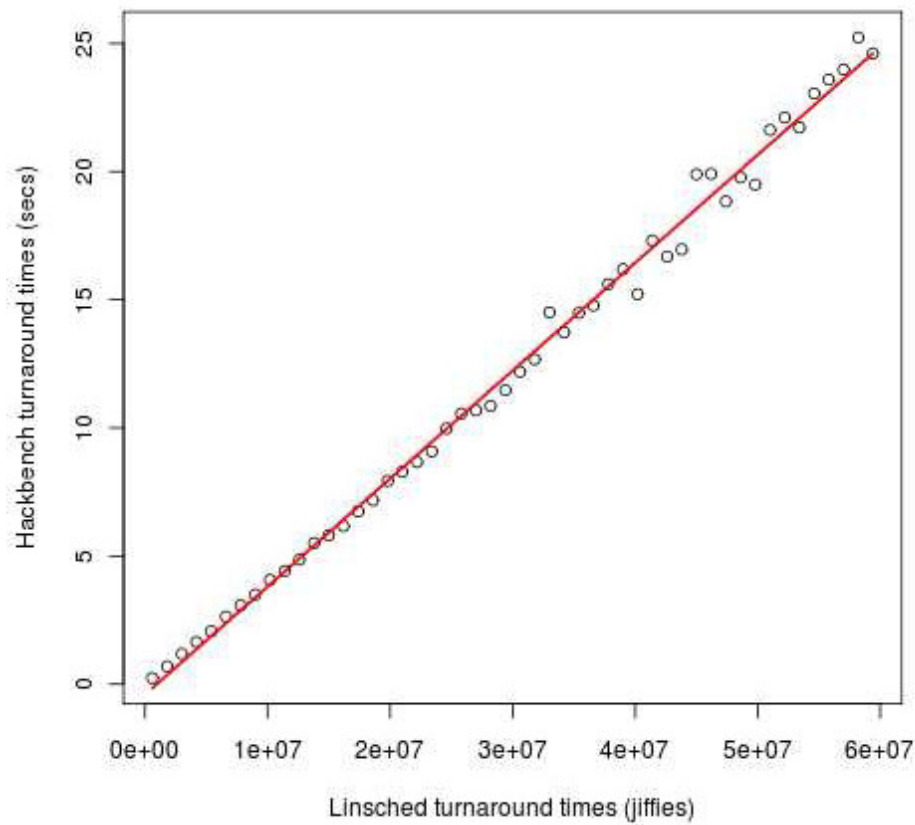

Fig. 1. Linear Relationship between LinSched Simulator Jiffies Metric and Hackbench Seconds Metric

We have ported the hackbench [9] scheduling benchmark to LinSched. This means that we have added modified hackbench code to the LinSched code. Therefore, the tasks LinSched creates are actually hackbench threads. The hackbench benchmark simulates a chat application with a group of 20 senders and 20 receivers. Each sender sends a message to each of the receivers, who receive from all the senders. The number of messages to be sent can be specified, as can the number of groups in the system. With our addition to LinSched, each sender is represented by a special "linsched writer callback" function and each receiver is represented by a special "linsched reader callback" function. These functions are implemented as finite state machines, so that if a sender is scheduled, for example, i.e. its callback is invoked, it continues where it left off last time it used the CPU. This means that it remembers its state. We have validated the simulator by running increasing sizes of hackbench workloads (5 groups of hackbench threads with varying number of messages to be sent by each sender). For each workload, we record the runtime when it runs on the LinSched simulator, and when it runs on a regular Linux installation. The relationship is linear, as is shown in Figure 1.

## V. OPTIMIZATION OF THE OPERATING SYSTEM SCHEDULER

Here we describe two approaches to tuning the Linux CFS scheduler. This scheduler has several parameters that can be set to tune its operation in order to achieve the best performance for a particular workload. We focus on three parameters, specified in *ns*. They are:

- Latency: A period of time during which all tasks should use a CPU. Range is 100, 000 to 1, 000, 000, 000*ns*.
- Minimum Granularity: The minimum period of time a task should have on a CPU. Range is 100, 000 to 1, 000, 000, 000*ns*.
- Wakeup Granularity: An additional allowance; a task is allowed to use a CPU for this amount of time, even after it has exhausted its primary allowance on the CPU. Rang is 0 to 1, 000, 000, 000*ns*.

### A. Global Optimization

We make use of the basic Particle Swarm Optimization algorithm, as described in [10], [11]. In this version, there are a set of particles, each of which has a certain location, *x* and a velocity *v*. The formula for updating a particle's velocity is:

$$\vec{v} \leftarrow w\vec{v} + \phi_p r_p (\vec{p} - \vec{x}) + \phi_g r_g (\vec{g} - \vec{x}) \qquad (1)$$

The formula for updating a particle's position is:

$$\vec{x} \leftarrow \vec{x} + \vec{v} \qquad (2)$$

The particles move through the search space updating, keeping track of their best position, as well as the swarm's best position. In equation (1) $w$ is called the inertia weight; $\phi_p$ is called the cognitive parameter and $\phi_g$ is called the social parameter. $\phi_p$ is set according to the weight the cognitive component should have. $\phi_g$ is set according the weight the social component should have. $r_p$ and $r_g$ are random numbers uniformly distributed in [0, 1]. $p$ is the best position of a particle and $g$ is the best global position.

We used meta-optimization [11] to come up with the $w$, $\phi_p$ and $\phi_g$ parameters. Our meta-optimization approach makes use of RSM [12], [13]. The tool we made use of is the *rsm* [14] package in the R statistical application [15]. We, first of all, designed an experiment for our three-parameter problem, using a Box-Behnken design. This resulted in a randomized experiment of 14 runs, with differing combinations of parameters. The $w$ parameter took on the values 0, 0.45, or 0.9. The $\phi_p$ and $\phi_g$ parameters took on the values 0, 2, or 4. Our PSO configuration had 20 particles and a maximum of 30 iterations. We came up with these numbers (20 and 30) through exploratory experiments using PSO together with hackbench (empirical search). These numbers allowed for sufficient variation in particle behavior before convergence. Table I shows the design of the experiments as well as the results. R* is the minimum response attained for that particular run i.e. the smallest turnaround time. I* is the number of iterations required to converge to the minimum response (the reader is reminded that our PSO algorithm uses 20 particles and a maximum of 30 iterations). $rsum = R^* + I^*$ is used to compare the various runs. The right-most three columns will be discussed in Section VI, the results section.

Table I
Meta-Optimization: RSM-Designed PSO Experiment Runs

| Run | $w$ | $\phi_p$ | $\phi_g$ | R* | I* | $rsum = R^* + I^*$ |
|---|---|---|---|---|---|---|
| 1 | 0 | 0 | 2 | 4294667318 | 5 | 4294667323 |
| 2 | 0.9 | 0 | 2 | 4294667318 | 4 | 4294667322 |
| 3 | 0 | 4 | 2 | 4294667318 | 29 | 4294667347 |
| 4 | 0.9 | 4 | 2 | 4294667318 | 30 | 4294667348 |
| 5 | 0 | 2 | 0 | 4294668048 | 31 | 4294668079 |
| 6 | 0.9 | 2 | 0 | 4294669150 | 31 | 4294669181 |
| 7 | 0 | 2 | 4 | 4294667315 | 7 | 4294667322 |
| 8 | 0.9 | 2 | 4 | 4294667315 | 6 | 4294667321 |
| 9 | 0.45 | 0 | 0 | 4294669038 | 31 | 4294669069 |
| 10 | 0.45 | 4 | 0 | 4294668498 | 31 | 4294668529 |
| 11 | 0.45 | 0 | 4 | 4294667315 | 5 | 4294667320 |
| 12 | 0.45 | 4 | 4 | 4294667315 | 5 | 4294667320 |
| 13 | 0.45 | 2 | 2 | 4294667315 | 18 | 4294667333 |
| 14 | 0.45 | 2 | 2 | 4294667315 | 16 | 4294667331 |

*B. Local Optimization*

We make use of Golden Section one dimensional optimization, as described in [10]. As discussed earlier, the Linux CFS scheduler has three parameters for tuning we are concerned with: latency, minimum granularity, and wakeup granularity. We name these *Latency*, *MinGran*, and *WakeupGran*, respectively. Within the CFS scheduling algorithm, another important variable is the *scheduling period* [5], [6]. This is actually the period of time during which all tasks should run once. We call this *SchedPeriod*. We name number latency, the maximum number of tasks that can be handled in *Latency*, *NrLatency*. The following equations are involved in calculation of the scheduling period.

$$NrLatency = \frac{Latency}{MinGran} \qquad (3)$$

If the number of tasks running,

$$NumTasks \leq NrLatency \quad (4)$$

then

$$SchedPeriod = Latency \quad (5)$$

Otherwise, we have the following:

$$SchedPeriod = Latency \times \frac{NumTasks}{NrLatency} \quad (6)$$

In order to apply the Golden Section, we convert the optimization problem from one depending on three variables, *Latency*, *MinGran*, and *WakeupGran*, into one depending on one variable, *SchedPeriod*. This is done by making useof the relationships between the various variables evident in the Linux kernel source code. So, instead of manipulating the values for the three variables, we manipulate only the *SchedPeriod* variable, and for each value of this variable, we discover the required values for *Latency*, *MinGran*, and *WakeupGran*, required to give this value. There are two possible discovery algorithms. Figure 2 shows the first Parameter Discovery Algorithm. This works for the case when *NumTasks* is greater than *NrLatency*. In other words, there are many tasks running. Here *Latency* is set at 19, 900, 000, the maximum required in order to use this approach. We know the number of tasks in the hackbench workload: 5 groups of 40 tasks = 200 tasks, since there are 20 senders and 20 receivers in each group.

```
if SchedPeriod > 200, 000, 000, 000 then
    WakeupGran ← SchedPeriod − 200, 000, 000, 000
    SchedPeriod = 200, 000, 000, 000
else
    WakeupGran ← 0
end if
Latency ← 19, 900, 000
MinGran ← SchedPeriod/200
return   Latency, MinGran, WakeupGran
```

Fig. 2. Parameter Discovery Algorithm 1

Figure 3 shows the second Parameter Discovery Algorithm. This works for the case when *NumTasks* is less than or equal to *NrLatency*. In other words, there are relativel few tasks running. Here *Latency* is the same as *SchedPeriod*. If the *Latency* is small (< 20, 000, 000), *MinGran* depends on *SchedPeriod* and *NumTasks*, otherwise it is set at 100, 000. The Golden Section search optimization method [10], [16] involves considering the *x* axis and two extreme points on this axis, $a_0$ and $b_0$. Two inner points ($a_1$ and $b_1$, to the right of $a_1$) are evaluated empirically, using the hackbench benchmark. If f ($a_1$) < f ($b_1$), the minimum lies in the range [$a_0$, $b_1$], otherwise it lies in the range [$a_1$, $b_0$]. A special constant, ρ, is used to divide the x axis into two segments, long and short. The Golden Section is the equation: ρ/(1 − ρ) = (1 − ρ)/1, i.e. ratio of short to long is equal to ratio of long to the sum of the two. The search was made more efficient by ensuring that the hackbench benchmark only runs once per iteration.

```
if SchedPeriod > 1,000,000,000 then
    WakeupGran ← SchedPeriod − 1,000,000,000
    SchedPeriod ← 1,000,000,000
else
    WakeupGran ← 0
end if
Latency ← SchedPeriod
if Latency < 20,000,000 then
    MinGran ← SchedPeriod/200
else
    MinGran ← 100,000
end if
return  Latency, MinGran, WakeupGran
```

Fig. 3. Parameter Discovery Algorithm 2

## VI. RESULTS

Partial results for meta-optimization of the PSO algorithm re given in table I and described in subsection A of Section V. *rsum* is used to compare the various runs (a combination of the minimum response achieved and the number of iterations required to achieve this). The minimum response is more important, by the summation. We fitted a second order model to these results and accepted the model because of high significance of the intercept, $w$ and $\phi_g$.

From our results, the value of $\phi_p$, the cognitive parameter did not have any effect on the speed of convergence or minimum response achieved of the PSO algorithm. The optimal parameter values were then obtained by use of the *rsm* package's *canonical path* optimization function. This makes use of gradients to find the steepest ascents in two directions. The values are: $w$ = 0.4365 and $\phi_g$ = 3.020. For completeness, $\phi_r$ was also set to 3.020.

We then ran the PSO algorithm using a hackbench workload on the LinSched simulator consisting of 5 groups of 40 processes each and 1 iteration of sending of messages. Results are shown in figures 4 and 5. Our optimized PSO produced the smallest turnaround time for the benchmark, 3028 Jiffies, while Golden Section algorithm 1 and 2 both produced Jiffies of 5939. However, it took Golden Section algorithm 1 just 8 iterations (benchmark executions) to reach its minimum, and Golden Section algorithm 2 9 iterations to reach its minimum, while it took the optimized PSO algorithm 31 benchmark executions to reach a minimum of 5939, equivalent to the Golden Section algorithms. From these results, it is clear that the optimized PSO algorithm yields the smallest turnaround time for our benchmark, but the Golden Section algorithms can yield useful results in a few iterations.

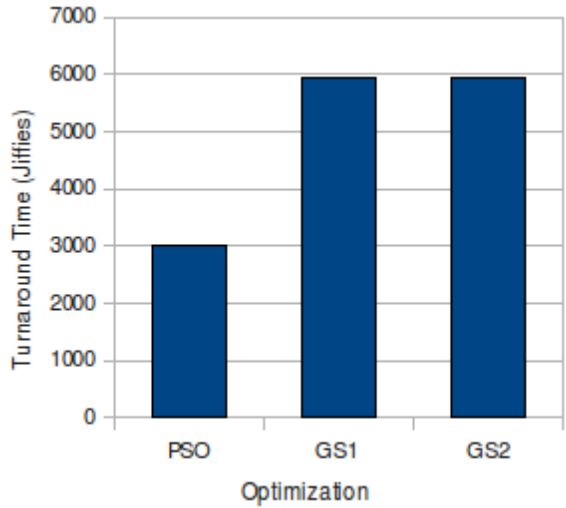

Fig. 4. Minimum Turnaround Times Achieved

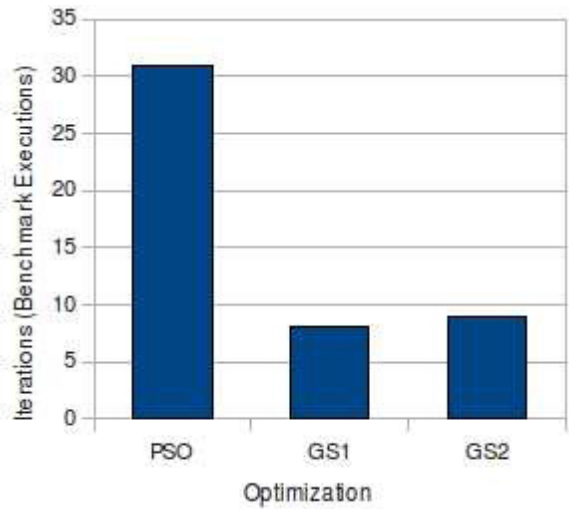

Fig. 5. Minimum Iterations Needed to Achieve Target Turnaround

VII. CONCLUSION

This paper has made several contributions. It has applied standard optimization methods to the task of operating system scheduler tuning. It has demonstrated the usefulness of Particle Swarm Optimization and used meta-optimization to come up with optimal parameters for the algorithm. It has also demonstrated how operating system kernel source code can be inspected in order to convert an optimization problem from one consisting of multiple design variables to one consisting of just one design variable. This conversion then makes it possible to make use of a local optimization approach, the Golden Section method. The optimized PSO algorithm yields the smallest turnaround time for our workload (the hackbench benchmark), but Golden Section yields useful results at a fraction of the time taken by PSO.